# DEFPOS Hα Observations of W80 Complex

Nazım Aksaker

Vocational School of Technical Sciences, University of Çukurova, Adana, Turkey;
*naksaker@cu.edu.tr*

**Abstract** We present Hα emission line measurements of the W80 nebular complex. A total of 26 regions have been observed inside the nebula with the Dual Etalon Fabry-Perot Optical Spectrometer (DEFPOS) system at the f/48 Coude focus of 150 cm RTT150 telescope located at TUBITAK National Observatory (TUG) in Antalya/Turkey. The intensities, the local standard of rest (LSR) velocities ($V_{LSR}$), heliocentric radial velocities ($V_{HEL}$) and the linewidths at Full Width at Half Maximum (FWHM) of the Hα emission lines have been determined from these observations. They lie in the range of 259 to 1159 Rayleigh ⋆ (R), 4 to 12 km s$^{-1}$ and 44 to 55 km s$^{-1}$, respectively. The radial velocity measurements show that there are several maxima and minima inside the W80. The new results confirm the literature that complex seems to be rather a uniform in radial velocity and no seen turbulent motion inside the complex. The average value of the calculated the Emission Measure (EM) for the regions is 3.1 pc cm$^{-6}$.

**Key words:** instrumentation: interferometers — ISM: H IIregions — techniques: radial velocities — techniques: interferometric

## 1 INTRODUCTION

The North America (NGC7000) and Pelican Nebulae (IC5070) complex together, catalogued as W80 (Westerhout 1958), is ∼ 3 east-southeast of Deneb in Cygnus. The region is located in the line of sight of the Cyg OB6 and OB7 associations which are part of the structure known as the Cygnus super bubble (Uyanıker et al. 2001). According to a recent study, an O5V star (2MASS J205551.25+435224.6) is located close to the geometric center of the complex (Comerón & Pasquali 2005). In the same paper, an assumed distance of 610 pc is given for this exciting source of W80. Straiys & Laugalys (2008) have identified a few more possibly highly reddened O-type stars that contribute to the ionization of the North America and Pelican nebulae. These two regions (NGC7000 and IC5070) are separated by a dust lane, L935 (Lynds 1962) which lies in the North-South direction.

---

⋆ 1R = $10^6/4\pi$ photons cm$^{-2}$ sr$^{-1}$ s$^{-1}$ = 2.4110$^{-7}$ erg cm$^{-2}$ sr$^{-1}$ s$^{-1}$ at Hα.

The W80 nebular complex, as a classical H II region, is very important for understanding the nature (physical structure, kinematics etc.) of such structures. Studying H II regions helps us to determine their sources of ionization (Aksaker et al. 2011) and to understand how these regions are related to the general structure of the ISM (Reynolds et al. 1998; Hausen et al. 2002). Therefore, observing H II regions in visual band using H$\alpha$ emission line gives valuable information (Aksaker et al. 2011; Fich et al. 1989). However, it is difficult to make accurate measurements of line fluxes by using traditional long-slit spectroscopic techniques, because of the low surface brightness and spatially extended emissions from large H II regions. Instead, high resolution spectral analysis of these faint, spatially extended sources requires high sensitivity and wide fields of view for spectrometers, examples of which are Dual Etalon Fabry Perot Optical Spectrometer - DEFPOS (Sahan et al. 2005, 2009; Aksaker et al. 2009) and Wisconsin H-Alpha Mapper - WHAM (Tufte 1997; Haffner et al. 2003)

The radial velocity with respect to the Local Standard of Rest ($V_{LSR}$), the Full Width at Half Maximum (FWHM), and the intensity of the H$\alpha$ emission line are obtained by DEFPOS observations very precisely. $V_{LSR}$ can be used for investigation of internal conditions in the mechanism such as the turbulence and inhomogeneities (Courtés et al. 1962; Williamson 1970). Measuring the FWHM of the H$\alpha$ and another emission line from a heavier atom (e.g. the $N^+$ or $S^+$), can be used to separate the thermal from the non-thermal motions in the gas (Reynolds 1988). Measured intensity of H$\alpha$ provides the emission measure (EM) value along the line of sight, which can be used in conjunction with the scattering measure, rotation measure, and dispersion measure to study interstellar turbulence. (Tufte et al. 1999; Haffner et al. 1998). These data help us to understand the nature of the H II regions and the Warm Ionized Medium (WIM) etc.

The layout of this paper will be as follows: In the first part, observations (Section 2.1) and data reduction (Section 2.2) are given. Then, $V_{LSR}$ for these regions are estimated in Section 3.1, the FWHM is estimated in Section 3.2 and its intensity in Section 3.3, respectively. Our conclusions and some suggestions for future works are given in the last section.

## 2 OBSERVATIONS AND DATA ANALYSIS

### 2.1 Observations

The observations were obtained by using the DEFPOS instrument on September 18-20, 2010 at TUBITAK National Observatory (TUG), Antalya, Turkey. The DEFPOS spectrometer measures the H$\alpha$ emission line covering a 200 km s$^{-1}$ (4.4 Å) spectral window with $\sim$ 30 km s$^{-1}$ spectral resolution within a field of view of size 4$'$. Detailed information about the instrument, optics, the methods of data analysis, and its intensity calibration can be found in our earlier work at Sahan et al. (2005), Sahan et al. (2009), Aksaker (2009), Aksaker et al. (2009).

A total of 26 regions were selected in W80 complex by examining the bright regions from the Digitized Sky Survey [1] (DSS). Thus, a high signal-to-noise ratio (SNR) was obtained in observations. Present observations covered the bright part of the W80 complex with 7 regions belonging to IC5070 and the rest to NGC7000. Exposure times were selected as 2400s to reach an optimum SNR. The journal of observations including the region ID numbers (col1), coordinates (col2&3), observation dates (col4), and exposure times (col5) are given in Table 1. The observations are also shown as colored (turquoise) circles on the DSS image of the region in Fig. 1.

---

[1] *http://skyview.gsfc.nasa.gov/*

**Table 1** Journal of Observations.

| Region Number | Coordinates | | Observation Date | Exposure Time |
|---|---|---|---|---|
| | $\alpha_{2000}$ | $\delta_{2000}$ | | |
| | (h:m:s) | (d:m:s) | | (s) |
| 1 | 20:58:10 | 43:24:59 | 18 Sep. 2010 | 2400 |
| 2 | 20:59:17 | 43:30:32 | ,, | ,, |
| 3 | 20:59:56 | 43:40:34 | ,, | ,, |
| 4 | 20:59:41 | 43:50:02 | ,, | ,, |
| 5 | 20:59:10 | 43:57:19 | ,, | ,, |
| 6 | 20:59:22 | 44:05:52 | ,, | ,, |
| 7 | 21:00:00 | 44:19:05 | ,, | ,, |
| 8 | 21:02:44 | 44:41:35 | ,, | ,, |
| 9 | 21:00:44 | 44:36:01 | ,, | ,, |
| 10 | 20:59:24 | 44:38:56 | ,, | ,, |
| 11 | 20:58:04 | 44:15:10 | ,, | ,, |
| 12 | 20:56:29 | 44:27:08 | 19 Sep. 2010 | ,, |
| 13 | 20:53:46 | 44:24:51 | ,, | ,, |
| 14 | 20:54:33 | 44:29:10 | ,, | ,, |
| 15 | 20:57:49 | 44:43:51 | ,, | ,, |
| 16 | 20:51:02 | 44:35:33 | ,, | ,, |
| 17 | 20:51:06 | 44:23:10 | ,, | ,, |
| 18 | 20:50:10 | 44:15:49 | 20 Sep. 2010 | ,, |
| 19 | 20:52:49 | 44:08:44 | ,, | ,, |
| 20 | 20:51:29 | 44:04:07 | ,, | ,, |
| 21 | 20:51:05 | 44:33:29 | ,, | ,, |
| 22 | 20:57:13 | 44:33:10 | ,, | ,, |
| 23 | 20:58:41 | 44:37:03 | ,, | ,, |
| 24 | 20:58:20 | 44:29:24 | ,, | ,, |
| 25 | 21:01:53 | 44:01:08 | ,, | ,, |
| 26 | 21:01:32 | 43:59:20 | ,, | ,, |

**2.2 Data reduction**

The observations were recorded by the 2048x2048 CCD camera mounted on RTT150 and DEFPOS as described in Sahan et al. (2009). The instrumental effects must be removed from the CCD resultant images before any data analysis. This process requires a series of steps including dark, bias, and flat field corrections, to remove the effects of cosmic rays, bad columns, and spurious reflections due to the DEFPOS optics. Reduced circular CCD images are then converted to one dimensional Hα spectra with 50 data points in each by the ring summing procedure (Coakley et al. 1996). A data reduction pipeline for DEFPOS spectra was available in IDL (Interactive Data Language) as described in Aksaker (2009) and Sahan et al. (2009). A sample reduced CCD image and its final spectrum are shown in Fig. 2.

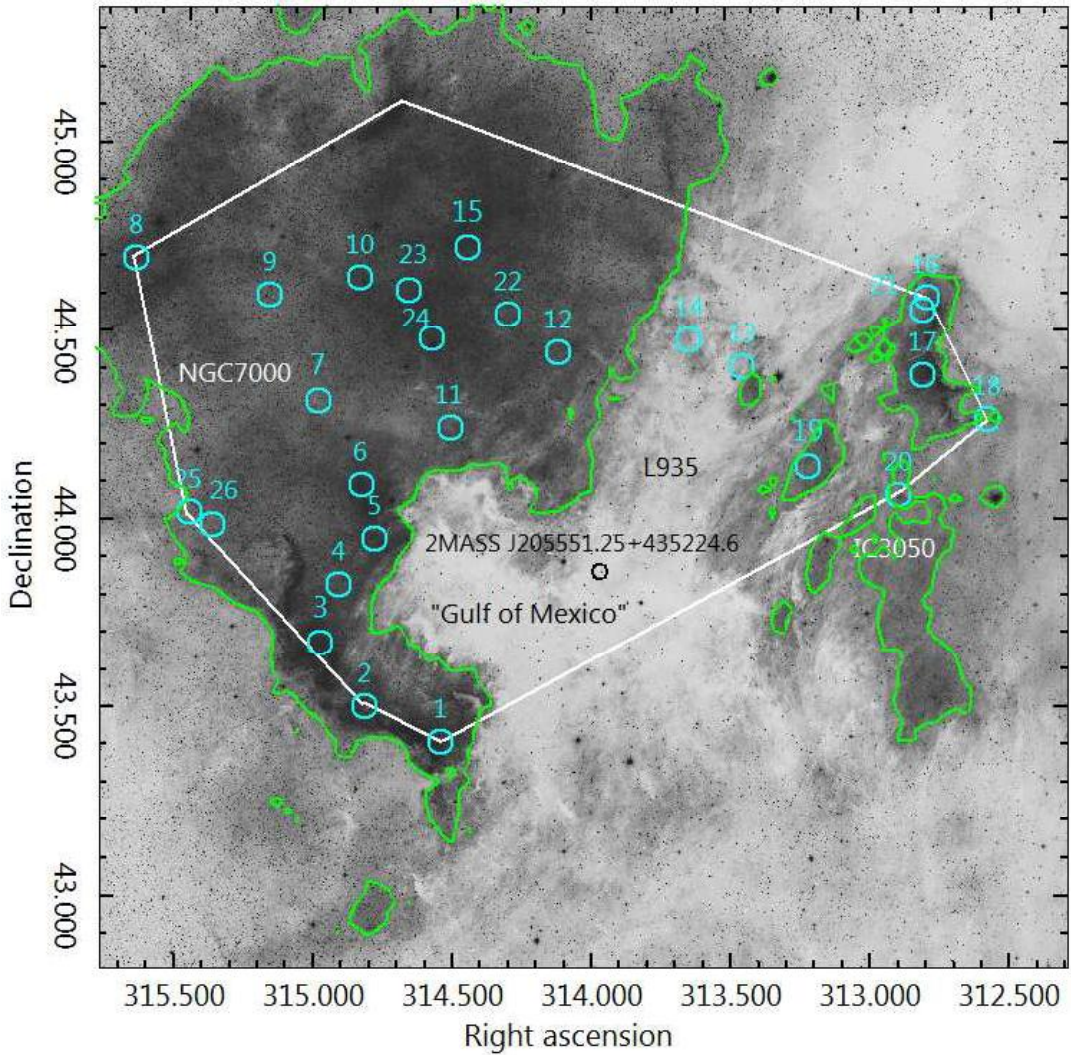

**Fig. 1** Optical image of the W80 region (DSS image) that comprising two HII regions: NGC 7000 (North America Nebula), IC 5070 (Pelican Nebula), and L935 (the dark lane). Circles (colored) show the relative size and positions of the observed regions. The black circle near the center marks the position of the exciting source proposed by Comerón & Pasquali (2005). The irregular rectangle in the investigating region shows the area modeled by DEFPOS data, corresponding to the field shown in Fig. 4. The green contour lines were obtained from brightness value of dss image to show borders of the nebulae. The coordinates are J2000 decimal degrees.

The SNR of a spectrum image can be calculated as the ratio of the amplitude of H$\alpha$ line to the amount of scatter in the residuals. The 2400 second integration time provides a SNR of $\simeq 20$ for a 900 R line. Details of the calculation of SNR in DEFPOS data can be found in Aksaker (2009).

By using a fitting program, the best Gaussian profile for each spectrum was then obtained and the best curve fitting for the data shown in Fig. 2 is typical and $r^2=0.984$, $r^2=1$ corresponding to a profile with a perfect fit. The intensity, the radial velocity, and the line width of the each H$\alpha$ spectrum was defined with the same procedure.

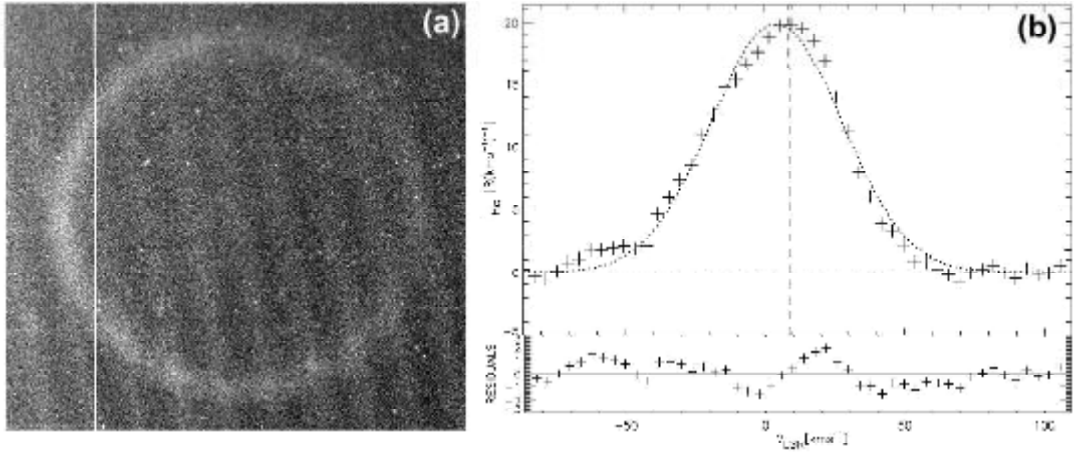

**Fig. 2** The brightest area was selected inside W80 as in Table 2. (a) A sample CCD image taken on 2010 September 19. (b) Top panel shows the spectrum and the residuals of the fit were plotted in the bottom panel. The spectrum consist of 50 spectral elements each corresponding to a 200 km s$^{-1}$ spectral window (a + symbol represents a 4 km s$^{-1}$ equivalently 0.087 Å wavelength ranges). The dashed horizontal bar defines the zero level (see the text for further details).

The real data points and their best fit spectral shape are shown with (+) symbols and the dotted line in Fig. 2b, respectively. In all Hα spectra, the horizontal axis is given in the velocity scale (km s$^{-1}$) and the vertical axis is the intensity in [R (km s$^{-1}$)$^{-1}$]. The radial velocities are with respect to the Local Standard of Rest (LSR) which is shown with a vertical dashed line at 8.9 km s$^{-1}$ in Fig. 2b. Velocity of this Hα data with respect to the LSR ($V_{LSR}$) is about 4.2±0.5 km s$^{-1}$. Its line width as FWHM is convolved to result a velocity value $V_{LSR}$ of 54.7±1.4 km s$^{-1}$. The corresponding intensity of Hα line is 1159.4±45.6 R.

The center of the geocoronal Hα is known to be shifted by - 2.33 km s$^{-1}$ from the rest wavelength of the recombination line at 6562.82 Å(Haffner et al. 2003; Aksaker et al. 2009). Thus, the LSR velocity values were corrected as $V_{LSR}$ = - $V'_{LSR}$ - 2.33 km s$^{-1}$. $V'_{LSR}$ came from velocity value of the best fit.

Instrumental line width of the DEFPOS itself was measured earlier to be 29.5 km s$^{-1}$ using a Thorium-Argon (Th-Ar) hollow cathode lamp. This value is very close to 30 km s$^{-1}$ the instrumental resolution of DEFPOS over the spectral window near the Hα. Therefore, the line width values are convolved by using a line fitting program and they intrinsic line widths are estimated by the quadratic subtraction of the 30 km s$^{-1}$ of instrumental line width (Aksaker et al. 2009).

In order for the intensity calibration for the DEFPOS data, a standard nebular source was used for each night before the observations. All calibrations were then tied to an absolute intensity measurement of the central part of the NGC 7000 at $\alpha = 20^h58^m04^s.0$, $\delta = +44°35'43''.0$ (equinox=2000.0) within a ∼ 4 arcmin spatial resolution. The Hα surface brightness at this point is about 900 R within the beam as measured by Aksaker et al. (2009).

## 3 RESULTS

The intensity ($I_{H\alpha}$ in R), the radial velocity (as $V_{LSR}$ and $V_{HEL}$), and the line width (as FWHM) of Hα emission lines measured and calculated in W80 Complex are given in Table 2 together with their ID numbers ($1^{st}$ column) and EM values are also given in a logarithmic scale (last column). $V_{HEL}$ values calculated with using the equatorial coordinates and $V_{LSR}$. Errors due to the scatter in the spectral data points have been determined by the standard deviation calculation carried out using a least-squares Gaussian fitting program. All of 26 DEFPOS spectra from W80 complex with their fits are in Fig. 3. All spectra are plotted in the same scale to see the differences clearly. The properties of the regions are further described below.

**Table 2** DEFPOS H$\alpha$ observations from W80 nebular complex.

| Region | Intensity (R) | FWHM (km s$^{-1}$) | $V_{LSR}$ (km s$^{-1}$) | $V_{HEL}$ (km s$^{-1}$) | $Log_{10}$(EM) (pc cm$^{-6}$) |
|---|---|---|---|---|---|
| 1 | 1035.7±50.6 | 52.9±1.7 | 8±0.6 | -8.0±0.6 | 3.4 |
| 2 | 953.1±43.3 | 53.1±1.5 | 10.7±0.5 | -5.2±0.5 | 3.3 |
| 3 | 1031.7±37.7 | 50.7±1.3 | 12.5±0.4 | -3.4±0.4 | 3.4 |
| 4 | 692.6±29.6 | 52.6±1.4 | 11.3±0.5 | -4.6±0.5 | 3.2 |
| 5 | 846.3±31.9 | 46.9±1.2 | 12±0.4 | -3.9±0.4 | 3.3 |
| 6 | 974.7±38.1 | 47.6±1.3 | 10.9±0.4 | -5.0±0.4 | 3.3 |
| 7 | 436.8±27.4 | 45.3±2.0 | 10.6±0.7 | -5.3±0.7 | 3.0 |
| 8 | 594.5±30.6 | 48.9±1.7 | 4.9±0.6 | -10.9±0.6 | 3.1 |
| 9 | 424.8±34.7 | 46.5±2.5 | 6.7±0.9 | -9.2±0.9 | 3.0 |
| 10 | 721.1±39.6 | 51.7±1.8 | 4.4±0.6 | -11.5±0.6 | 3.2 |
| 11 | 586.7±35.4 | 49.0±1.9 | 9.3±0.7 | -6.7±0.7 | 3.1 |
| 12 | 813±34 | 51.7±1.4 | 8±0.5 | -8.0±0.5 | 3.3 |
| 13 | 342.9±32 | 47.4±3.6 | 5.9±1.1 | -10.2±1.1 | 2.9 |
| 14 | 259.8±26.5 | 45.7±3.1 | 5.5±1.1 | -10.6±1.1 | 2.8 |
| 15 | 1159.4±45.6 | 54.7±1.4 | 4.2±0.5 | -11.8±0.5 | 3.4 |
| 16 | 537.7±34.5 | 50.7±2.1 | 7±0.7 | -9.2±0.7 | 3.1 |
| 17 | 850.7±40.1 | 55.5±1.6 | 5.6±0.6 | -10.6±0.6 | 3.3 |
| 18 | 324.1±34.1 | 52.2±3.7 | 11.8±1.3 | -4.5±1.3 | 2.9 |
| 19 | 447.8±49.9 | 48.8±4.0 | 8.9±1.3 | -7.3±1.3 | 3.0 |
| 20 | 360.6±25.3 | 48.2±2.2 | 8.3±0.8 | -7.9±0.8 | 2.9 |
| 21 | 472±25.1 | 48.9±1.7 | 6.2±0.6 | -10.0±0.6 | 3.0 |
| 22 | 678.3±32.5 | 49.9±1.5 | 8.7±0.6 | -7.3±0.6 | 3.2 |
| 23 | 764.3±39.2 | 48.6±1.6 | 5.9±0.6 | -10.0±0.6 | 3.2 |
| 24 | 1019.9±47.4 | 61.5±1.8 | 11.2±0.6 | -4.8±0.6 | 3.4 |
| 25 | 290.2±29.8 | 46.8±3.1 | 5.2±1.1 | -10.6±1.1 | 2.8 |
| 26 | 351.2±29.3 | 44.3±2.4 | 8.5±0.9 | -7.3±0.9 | 2.9 |

To complete the analysis, we also constructed 2 dimensional maps of intensity, $V_{LSR}$, and FWHM using the data by interpolation techniques. Corresponding maps are given in Fig. 4. The maps were obtained with the help of TRIANGULATE and TRIGRID functions written in IDL using the observations.

### 3.1 The $V_{LSR}$

All the radial velocities presented are with respect to the LSR. The systematic uncertainties in the data are typically of 2-3 km s$^{-1}$. Contribution to this uncertainty from the velocity calibration from random noise is less than 2 km s$^{-1}$ (see Table 2), same as in previous works (Aksaker et al. 2009; Aksaker 2009). The average radial velocity ($V_{LSR}$) is 8.5±0.6 km s$^{-1}$ for the NGC7000, 7.4±0.9 km s$^{-1}$ for the IC5070 and 8.2±0.7 km s$^{-1}$ for the entire W80 complex. Also, the average radial velocity ($V_{HEL}$) is -7.4±0.6 km s$^{-1}$ for the NGC7000, -8.8±0.9 km s$^{-1}$ for the IC5070 and -7.8±0.7 km s$^{-1}$ for the entire W80 complex.

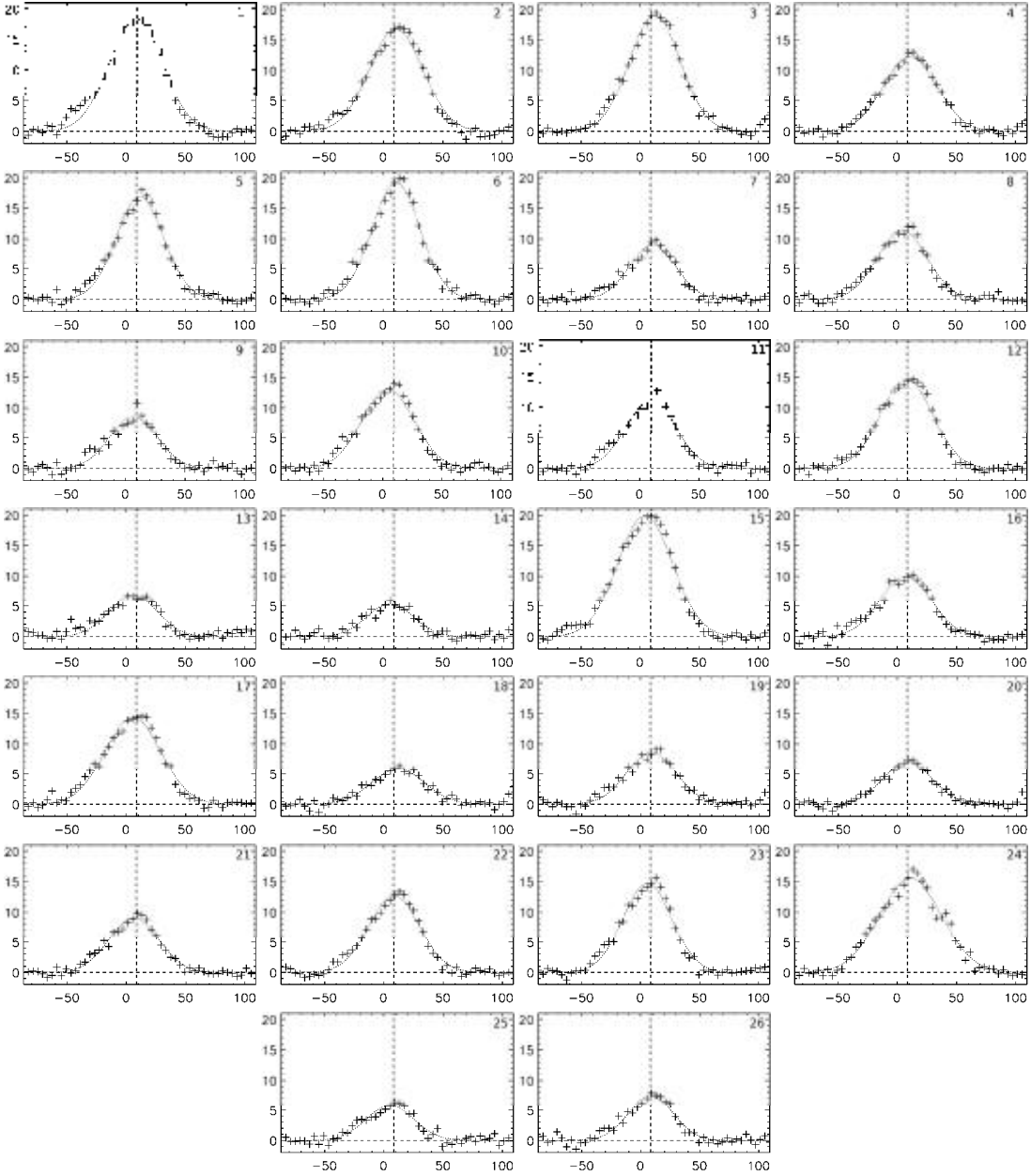

**Fig. 3** DEFPOS spectra (velocity [km s$^{-1}$] vs intensity [R (km s$^{-1}$)$^{-1}$]) of the W80 nebular complex, with their ID numbers as in Table 1 and 2, shown at their upper right corners.

One can accept that these values could be taken as the velocity of NGC7000 with respect to LSR. In general, inside of both NGC7000 and IC5070 complexes, the radial velocities decrease towards the north. There are several maxima and minima inside the NGC7000. The radial velocity of periphery of "Gulf of Mexico" region is almost the same value with the average radial velocity. The fastest radial velocity is 12.5±0.4 km s$^{-1}$ (region 3) at the south of the NGC7000, near the Gulf of Mexico, and the slowest is 4.2±0.5km s$^{-1}$ at the north of the NGC7000. Average values for NGC7000 as 8.5±0.6 km s$^{-1}$ and for IC5070 as 7.4±0.9 km s$^{-1}$ are quite near to each other and differences between NGC7000 and IC5070 is less than the uncertainties involved. We can conclude that the W80 seems to have rather uniform radial velocity distribution.

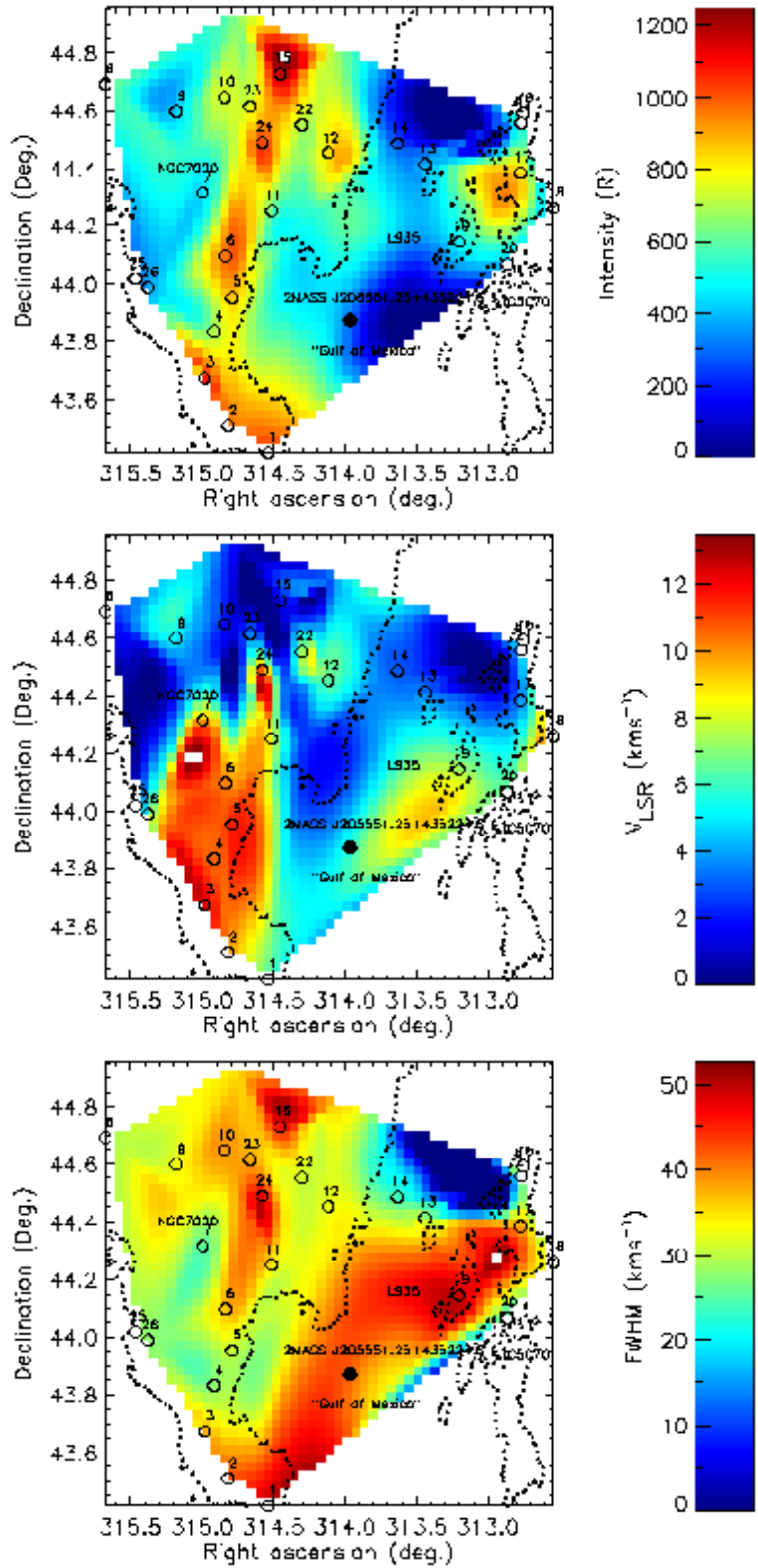

**Fig. 4** Interpolation maps of W80 complex using the present Intensity, $V_{LSR}$, and FWHM data. The areas identifying IC 5070, L935, NGC 7000, observed regions, and dotted contour lines are in the same locations as Fig. 1

### 3.2 The FWHM

The systematic uncertainty at FWHM of the Hα is always less than the size of each data point (<4 km s$^{-1}$) in the spectra. The random noise in the data is similar to the radial velocity values. The average FWHM is 50.2±1.8 km s$^{-1}$ for the NGC7000 and 49.7±2.8 km s$^{-1}$ for the IC5070 and 50.0±2.1 km s$^{-1}$ for the total W80 complex. The central parts of the NGC7000 and IC5070 have a wider FWHM than the rest of the regions. This could be a clue for lower temperature values inside. However, we do not have any information about the emission lines of heavier elements which may further substantiate this conclusion.

### 3.3 The Intensity

There was approximately 15% uncertainty in the absolute intensity calibration plus a ∼ 9% uncertainty emerging from the sources of random photons of the data obtained from W80. The average intensity is 743±36 R for the NGC7000 and 449±33 R for the IC5070 and 653±36 R for the W80 complex. A bright line can be seen in Fig. 4 which lie from north to south of the NGC7000. The night sky brightness is about 250 R. Thus, Both NGC7000 and IC5070 can be seen in the night sky with a naked eye.

We also calculated EM (in the absence of extinction) from the DEFPOS data integrated over velocities $V_{LSR}$. < 100 km s$^{-1}$ using EM = $\int n_e^2$ dl = 2.75$T_4^{0.9} I_{H\alpha}$ (pc cm$^{-6}$) where $n_e$ is the local electron density, dl is distance element to the source region, $I_{H\alpha}$ is the Hα intensity in R, $T_4$ is the temperature of the gas in units of $10^4$K (Haffner et al. 2003; Reynolds et al. 2005). A temperature of about 8000 K for the both regions (NGC7000 and IC5070) is consistent with the widths of the Hα, [O I], and [S II] emission lines (e.g., Reynolds 1988). The above calculation of the EM values are in line with above calculations and are given in Table 2. The average EM is about 3.1 (pc cm$^{-6}$) for the entire W80.

## 4 COMPARISON WITH OTHER WORKS

### 4.1 The WHAM

One another currently in use Fabry-Perot spectrometer is the WHAM (Wisconsin H-Alpha Mapper). WHAM observation offers within a 200 km s$^{-1}$ (about 4Å near Hα) spectral window with 8-12 km s$^{-1}$ velocity resolution from a one-degree, spatially integrated beam of the sky (Haffner et al. 2003). With its large-aperture (1 degree field of view) design and modern CCD technology, WHAM can detect emission as faint as 0.05 R in a 30 second exposure. The calibrated spectra and velocity interval maps are available and can be downloaded from the WHAM website (www.wham.astro.wisc.edu). The wham data cover the W80 region with a 5 data point which have the galactic coordinates, Intensity, $V_{LSR}$ and FWHM values listed in Table 3

For the WHAM data of the W80 region, means of the Intensity, the $V_{LSR}$ and the FWHM results were 264.3 R, 2.6 km s$^{-1}$ and 33.2 km s$^{-1}$, respectively. Comparing WHAM data and DEFPOS data is not the right approach because of the different FOV. However it can be seen that the $V_{LSR}$ and the FWHM values of both instruments are in agreement.

**Table 3** The WHAM data near the W80 nebular complex.

| $l$ (deg) | $b$ (deg) | Intensity (R) | $V_{LSR}$ (km s$^{-1}$) | FWHM (km s$^{-1}$) |
|---|---|---|---|---|
| 84.2820 | -0.0027 | 135.7 | 4.5 | 44.8 |
| 84.7820 | 0.8475 | 73.2 | 1.0 | 44.2 |
| 85.2621 | -0.0027 | 189.1 | 3.3 | 35.3 |
| 85.3014 | -1.7008 | 483.3 | 0.8 | 40.0 |
| 85.7614 | -0.8508 | 311.4 | 5.1 | 33.2 |

### 4.2 Fountain et al. (1983)

The multiple slit echelle spectrograph observations of the H$\alpha$ were used to map the radial velocities of the W80 complex by Fountain et al. (1983)(hereafter F83). The instrument has 2 arcmin x 3 arcmin FOV, 12.7 cm aperture and dispersion resolution of 7.25 Å mm$^{-1}$. They measured the radial velocity and FWHM of the H$\alpha$ line of the W80 complex. The mean value of the $V_{HEL}$ of the complex -15.1±5.5 km s$^{-1}$ and the FWHM mean value is 28.6±0.6 km s$^{-1}$. Although, we have achieved to similar results of F83 but not close agreement because of field of view, number of data and their coverage. They were also reported that no significant variation in radial velocity.

## 5 DISCUSSION AND CONCLUSIONS

There are many high spatially resolved and flux calibrated of H$\alpha$ images surveys for northern sky such as the INT Photometric H$\alpha$ Survey of the Northern Galactic Plane (IPHAS; Drew et al. 2005) and the Virginia Tech Spectral-Line Survey (VTSS; Dennison et al. 1998). Although their data have better spatial resolution. They have not include velocity information. For this reason, they cannot provide alone sufficient information to understand H IIregions. At this point high spectral and spatial resolution Fabry-Perot spectrometers such as scanning Fabry-Perot (Godbout et al. 1998), Integral Field Spectroscopy (IFS; e.g, Jahnke et al. 2004) and echelle spectrometers (F83) can show small-scale velocity fluctuations (turbulent motion) inside H IIregions. This advantage comes with a challenge of long exposure time and huge data.

According to the above consideration the DEFPOS instrument seats in the middle of the both techniques. It can measure the intensity (I$_H\alpha$), the line width (FWHM) and the radial velocity ($V_{LSR}$) of H$\alpha$ precisely. Such spectrometer is a unique tool to investigating H$\alpha$ emission line from diffuse sources like H IIregions.

The results of the DEFPOS data from W80 the intensity (I$_H\alpha$), the line width (FWHM) and the radial velocity ($V_{LSR}$) of H$\alpha$ are in the ranges of 259 to 1159 Rayleigh, 4 to 12 km s$^{-1}$ and 44 to 55 km s$^{-1}$, respectively (see Table 2). Their respective average values are 653±36 R, 8.2±0.7 km s$^{-1}$ and 50.0±2.1 km s$^{-1}$.

The average radial velocity ($V_{HEL}$) is -7.4±0.6 km s$^{-1}$ for the NGC7000, -8.8±0.9 km s$^{-1}$ for the IC5070 and -7.8±0.7 km s$^{-1}$ for the entire W80 complex. F83 found that the mean values of the $V_{HEL}$ is -15.4±3.7 km s$^{-1}$ for the NGC7000, -14.8±4.1 km s$^{-1}$ for the IC5070 and -15.1±5.5 km s$^{-1}$ for the entire W80 complex. Also, Hippelein (1973) gave heliocentric radial velocity of -15.7 and -16.4 km s$^{-1}$ for NGC7000 and IC5070, respectively.

The DEFPOS spectrometer have ability of precise measurements such as the radial velocity ($V_{LSR}$) of Hα. However it cannot obtain the detailed turbulent motion inside the nebulae because of the relatively wide FOV. Its radial velocity data can utilize to obtain a presence of large-scale velocity gradients. This can be seen apparently in the radial velocity map of the W80 complex (see Fig. 4). Results of earlier works produced a general picture of the dynamics of the W80 complex. There is no turbulence in the W80 complex, have been identified. Additionally we confirm their results with the new DEFPOS data.

There are several maxima and minima inside the NGC7000 and IC5070. The average EM value calculated, using (R) of Hα, is 3.1 (pc cm$^{-6}$) for the entire W80.

It seems from the profile plots in Fig. 3 for region 1-6 which show bright lines in both $I_{H\alpha}$ and $V_{LSR}$. Moreover, FWHM values from same regions also wider. This region shows a circular ionized nebula around an early type star (2MASS J205551.25+435224.6) like classical H IIregions. That could be proven with the help of more data in other wavelength.

The 2400s integration time per beam provided a SNR of 5-20 for observed 260-1160 R line intensity having widths of 46-55 km s$^{-1}$. For comparison, note that 30s integration time per beam provides a SNR of 20 for a weak 0.5 R line having a width of 20 km s$^{-1}$ (see Tufte (1997)). Using different equipment our measurements seem to have quite low SNR values. However, these values are good enough to work large and weak diffuse objects.

There is an expected correlation with the $I_{H\alpha}$ and the FWHM values of the Hα line. However, the same could not be established neither between the FWHM and the $V_{LSR}$ nor between $I_{H\alpha}$ and the $V_{LSR}$ values. By repeated observations, changes in physical properties over time could be obtained. This is one of our future aims.

**Acknowledgements** This work is supported by the TUBITAK with grant number 104T252 and also by Scientific Research Project office of Çukurova University with grant number TBMYO2010BAP4. The author thank to TUBITAK(The Scientific and Technical Research Council of Turkey) for a (partial) support in using RTT150 (project number 09ARTT150-436-1 and 11ARTT150-116-1) and TUBITAK National Observatory (TUG) staff for their help. The author would like to thank to Ilhami Yegingil and Muhittin Sahan for their valuable guidance, help and advices. The author also would like to thank R. J. Reynolds (University of Wisconsin) for his valuable help in the optical design of the DEFPOS as well as to start present study. The author also grateful to A. Akyuz, M. E. Ozel and A. Abdulvahitoglu for reading and correcting the manuscript and for their valuable remarks.